\documentclass{kapproc} 
\newcommand{\beq}{\begin{equation}}
\newcommand{\eeq}{\end{equation}}
\newcommand{\cs}{c_{\rm s}}
\newcommand{\cmc}{{\rm cm}^{-3}}
\newcommand{\cms}{{\rm cm}^{-2}}

\RequirePackage{graphicx}%
\RequirePackage{epsf}%
\input{psfig.sty}

\upperandlowercase
\let\footnote\savefootnote
\let\footnotetext\savefootnotetext 
\let\footnoterule\savefootnoterule 
\kluwerbib 

\begin{document}


\articletitle{Turbulence and Magnetic Fields in Clouds}

\articlesubtitle{Discussion}

\chaptitlerunninghead{Turbulence and Magnetic Fields}



\author{Shantanu Basu}
\affil{Department of Physics and Astronomy, The University of Western Ontario,
\\ London, Ontario N6A 3K7, Canada}
\email{basu@astro.uwo.ca}





\begin{abstract}
We discuss several categories of models which may explain the IMF,
including the possible role of turbulence and magnetic fields.
\end{abstract}

\section{Introduction}
Given the pervasive presence of non-thermal motions in molecular
clouds and evidence for energetically significant magnetic fields,
it is tempting to suggest that turbulence and/or
magnetic fields play a critical role in determining the stellar
initial mass function (IMF). In this discussion, we 
review several categories of IMF models, and discuss how they are
influenced by turbulence and magnetic fields.

On one hand, the IMF can be thought to be determined by a 
direct mapping from the core (or condensation) mass function (CMF),
if the core truly represents a finite mass reservoir for star formation.
Alternatively, the IMF may be determined from interactions that happen 
very close to a forming protostar, as it
accretes matter from its parent core. In the latter case, the 
CMF may not be directly mapped onto the IMF.
We review several possibilities in the next two sections.

\section{The CMF leads to the IMF}

The main difficulty to overcome here is the definition of a core
itself. A core boundary is not nearly as 
well defined as a stellar surface, so the mapping of a
CMF to IMF is problematic from the outset. 

Cores have often been defined as a region within which emission
from a certain molecule is detected. This is hardly a physical
demarcation. More recently, near-infrared absorption maps 
(Bacmann et al. 2000) have captured
the merger of a density profile into the background. This may
represent a physical boundary. A theoretical definition of a core
boundary may rely on the presence of a magnetically subcritical
envelope around a supercritical inner region (the core), or it may
rely on the (usually larger) gravitational zone of influence
of a density peak.

In any case, there are three main candidates for the determination
of the CMF: (1) pure gravitational 
fragmentation; (2) turbulent fragmentation, and (3) magnetically
regulated fragmentation. Furthermore, in an extension to these models,
the CMF (if clearly definable) may develop a power-law tail 
due to accretion effects. We treat this as a fourth possibility
which is not independent of the first three.


\subsection{Gravitational fragmentation}

Any non-isotropic medium that is dominated by gravity is expected to 
fragment into Jeans-mass type fragments, through an initial 
collapse into a sheet, followed by the break up of the sheet.
This is the famous Zeldovich (1970) hypothesis in cosmology. 
In the interstellar medium, sheet-like initial configurations may 
be promoted by effectively one-dimensional
compressions due to supernova shock waves and expanding HII regions,
or by relaxation along magnetic field lines.
The preferred fragmentation scale in an isothermal non-magnetic
flattened sheet of
column density $\Sigma$ is $\lambda_{\rm m} = 4.4 \, \cs^2/(G \Sigma)$
(Simon 1965), where $\cs$ is the isothermal sound speed and
$G$ is the gravitational constant. The formation time for a 
cluster of stars is effectively the sound crossing time
across this distance, typically less than a few Myr for most
molecular clouds. Mass that does not accrete to one gravitating
center is within the gravitational
sphere of influence of a neighboring core.
We note that the resulting CMF (by any definition) from this kind of 
fragmentation will likely be peaked around the Jeans mass
$M \sim \Sigma \lambda_{\rm m}^2$, but has not yet been calculated in
detail.

The most likely candidates for gravitational fragmentation are the embedded
clusters in which multiple stars are forming in close proximity,
and which account for a majority of star formation (Lada \& Lada 2003).
However, the above authors also point out that the
star formation efficiency (SFE, defined as the fraction of gas mass 
converted into stars) in these clusters is still
quite low, in the 10\% - 30\% range. Perhaps the feedback effect
of outflows can explain at least the upper values ($\sim$ 30\%) 
of SFE's (e.g., Matzner \& McKee 2000). A full numerical simulation of
the feedback on a cloud from outflows is still prohibitive.
However, pure gravitational fragmentation does seem to be excluded as a 
possibility for giant molecular clouds (GMC's) as a whole, since 
their overall SFE is only a few percent (Lada \& Lada 2003). 
This point can be traced back to Zuckerman \& Evans (1974).





\subsection{Turbulent fragmentation}
A way to explain the low SFE is to postulate that turbulent support
prevents gravitational fragmentation on large scales, but that cores are 
also created by turbulent compressions. 
This is broadly consistent with the observation
that turbulent motions dominate on large scales but become sub-thermal
on dense core scales, in accordance with the well-known linewidth 
($\sigma$)-size ($R$) relation, $\sigma \propto R^{0.5}$ (e.g., 
Solomon et al. 1987). Strong turbulent driving in clouds can explain 
the overall low SFE (see Vazquez-Semadeni, this volume), by keeping
most material in a disturbed and non-self-gravitating state.
It has also been shown that isothermal turbulence
leads to a lognormal probability density function (pdf) for the 
gas density (e.g., Padoan, Nordlund, \& Jones 1997; Passot \& 
Vazquez-Semadeni 1998; Scalo et al. 1998;
Ostriker, Stone, \& Gammie 2001; Klessen 2001).
Elmegreen (2002) demonstrates that a lognormal density
pdf will lead to power-law clump mass distribution when thresholded
at various levels, with different indices for different threshold levels.
Padoan \& Nordlund (2002, see also Padoan in this volume) 
also demonstrate that
a lognormal density pdf is consistent with a power-law CMF given 
that the power-spectrum is a power law, and assuming that the 
cores have sizes comparable to the thickness of post-shock gas layers.

In all models of turbulent fragmentation, an important question arises:
is the CMF just a property of
how cores are defined, or does it represent the finite reservoirs
of mass that may be available for star formation?
A further problem is that turbulence tends to decay away in a 
crossing time if not continually driven, so that turbulent fragmentation
may quickly give way to gravitational 
fragmentation.
If the latter leads to runaway peaks and star formation within a 
crossing time,
we are again hard pressed to understand the overall low SFE of GMC's.

\subsection{Magnetically regulated fragmentation}

Real interstellar clouds are both turbulent and contain 
magnetic fields which are in approximate equipartition
with gravity (e.g., Crutcher 1999). The turbulence itself likely consists of
MHD disturbances. Therefore, a realistic scenario is that
of turbulent dissipation followed by magnetically regulated fragmentation
in dense regions. The unique features of fragmentation of clouds with
near-critical mass-to-flux ratio are: (1) a longer timescale for collapse
than simply the hydrodynamic crossing time, and (2) the outer envelopes
may remain supported against global collapse. 
Basu \& Mouschovias (1995) have demonstrated that a magnetically
supercritical fragment within a subcritical envelope evolves very
rapidly once it is large enough to also be thermally unstable. 
The resulting collapse scale is smaller than the 
the original fragmentation scale of a subcritical cloud.
Hence, an inter-core medium exists which is subcritical and remains in
a state of slow evolution.
Furthermore, Basu \& Ciolek (2004)
have demonstrated that even if the background cloud has an exactly 
critical mass-to-flux ratio, the mass and flux redistribution effected
by ambipolar diffusion naturally leads to both supercritical cores 
and a subcritical envelope. 

Magnetic fields may also prevent unstable fragments from becoming
extremely elongated, as occurs in models of pure gravitational fragmentation
(Miyama, Narita, \& Hayashi 1987). Two-dimensional magnetic
fragmentation models of Basu \& Ciolek (2004) show much
milder elongations when magnetic fields are significant, and are
in principle more 
consistent with the inference from observations that cores are 
overall triaxial 
but more nearly oblate than prolate (Jones, Basu, \& Dubinski 2001).
Their results also show that the magnitude of infall motions and 
the preferred fragmentation scale are dependent on the initial
mass-to-flux ratio.
Li \& Nakamura (2004, also Nakamura this volume) have developed a 
model of turbulent fragmentation in a subcritical cloud with 
ambipolar diffusion, also using a two-dimensional simulation.
They show that supercritical fragments can be formed in a few
Myr, but that the magnetic field helps maintain a relatively low SFE.

A key challenge to this theory is to find the putative subcritical
envelopes through highly sensitive Zeeman observations of molecular
cloud envelopes. If subcritical envelopes are observed, this will 
go a long way toward explaining the low inferred SFE. Current magnetic
field data is consistent with a near-uniform (and near-critical)
mass-to-flux ratio 
in the column density range $10^{21} \, \cms - 10^{23} \, \cms$. However,
there is an apparent mild bias toward subcritical mass-to-flux ratios at low
column densities (Crutcher 2004).

\subsection{Accretion modification of the CMF}

If any of the above scenarios lead to a lognormal initial
distribution of core masses, it is possible that the gravitational
influence of the core on the surrounding cloud will lead to accretion
that alters the distribution of masses $M$. An original model of this type
is due to Zinnecker (1982), in which Bondi accretion 
($dM/dt \propto M^2$) leads to a 
power-law tail in the number ($N$) of stars per unit mass interval, 
i.e., $dN/dM \propto M^{-2}$. This is 
due to larger initial masses growing at a relatively faster rate.
A different model in this category is due to Basu \& Jones (2004,
see also this volume);
they show that an exponential distribution of accretion lifetimes 
and accretion rate $dM/dt \propto M$ 
can lead to a power-law tail in $dN/dM$.
This can produce an IMF with a lognormal body and a power-law tail.
A similar explanation was offered by Myers (2000).

\section{The IMF from star-core interactions}

There are two main ideas for how the IMF may be determined by interactions
occurring very close to a forming protostar: (1) outflow limited
accretion, and (2) termination of accretion by an ejection process.
In an extreme form of this approach, the CMF is irrelevant because
infall is terminated before any finiteness of the available
mass comes into play.

\subsection{Outflow interactions}

Strong outflows are present in the earliest observed stages of 
protostellar accretion (see Andr\'e, this volume). It has
been proposed that winds and/or swept-up outflows can reverse infall 
(e.g., Shu, Adams, \& Lizano 1987).
Some IMF models have been developed based on this concept. 
Adams \& Fatuzzo (1996, see also Adams in this volume)
have argued that 
mass accretion will be halted when its rate drops below
the mass outflow rate. The presence of a variety of multiplicative 
input parameters leads them to infer 
a near-lognormal distribution for the IMF. 
A similar model has been presented by Silk (1995, see also this volume), 
which results in a power-law IMF.
However, it is still not
clear that outflows are sufficiently wide-angled to reverse
infall in all directions, so a finite mass reservoir may still
be necessary.

The best scenario may be a combination of a nearly finite
mass reservoir and the action of outflows to clear residual material.
Shu, Li, \& Allen (2004) have carried forward the type of model presented
by e.g., Basu \& Mouschovias (1995) to its logical limit, by studying
the accretion onto a protostar from a subcritical envelope.
They find that a final equilibrium is possible only if the 
gravity of the point mass (the protostar) at the core
can be offset by unrealistically
large amounts of magnetic flux within the protostar.
The breakdown of ideal MHD near the protostar will ultimately 
prevent magnetic levitation of the subcritical envelope, but outflows
are invoked as a last line of defense against the low-level infall
from the subcritical envelope.

The above scenario may be appropriate to explain isolated star
formation as well as cluster formation in which the SFE is 
quite low. For regions that give rise to bound open clusters
(a distinctively rare occurrence according to Lada \& Lada 2003)
the SFE must be rather high for the cluster to remain bound.
In such cases, 
a simple gravitational (or highly supercritical) fragmentation model may 
be adequate.


\subsection{Competitive accretion}

Another process that occurs deep within a core is the interaction
between protostars that may have formed in the same core. Multiple
protostars come from direct fragmentation during collapse, or from the
fragmentation of a circumstellar disk after the first protostar has
formed. Bate, Bonnell, \& Bromm (2003, see also Bate in this volume) 
have argued that the IMF can
be explained by the evolution of multiple 
protostars which start out with a mass approximately equal to 
the Jeans mass for the 
density $n \sim 10^{10} \, \cmc$ at which the gas becomes opaque.
Dynamical interactions between the protostars then cause
them to be ejected from the parent core at various stages of
mass accretion from that core.
Their simulations show this effect and the calculated protostar mass 
distributions resemble a lognormal, but may
be interpreted as having a weak power-law tail.
In this picture, star formation is very efficient, and the problem
of low SFE is pushed back to the unmodeled regions outside the 
cluster-forming cores. Turbulence and magnetic fields are also not
required except to understand the outer unmodeled regions.

\section{Conclusions}

Both turbulence and magnetic field effects are important physical
processes in molecular cloud evolution, and are a great challenge to 
theorists due to the complexity of the nonlinear equations that 
describe them. However, for the purpose of this discussion, it
is also worth asking the hard question: do these effects fundamentally
affect the IMF?
Heyer (this volume) has questioned the existence of any linkage 
between turbulent properties of a cloud and the rate of star
formation within them.
Large-scale magnetic fields are also 
invoked as a formidable opponent to gravity, but if most stars form
in local cluster-forming regions which are supercritical, then the
magnetic field may not be a dominant effect.
On the other hand, magnetic fields are necessary for the 
generation of outflows, which are in turn invoked to explain why observed
dense embedded clusters have an SFE no greater than about 30\%.
A key outstanding question is whether outflows can really limit the 
SFE to 30\% (or less!) in a highly supercritical cloud region.
An ultimate model of star formation will likely have to 
account for the low SFE of GMC envelopes (using turbulent and/or
magnetic effects) as well as include the
self-consistent feedback effect of outflows upon gravitational collapse.


\begin{acknowledgments}
I wish to thank the organizing committee for 
the outstanding concept of IMF@50 and for creating a 
remarkably stimulating meeting. 
I also thank Martin Houde, Carol Jones, and Takahiro Kudoh
for their comments on the manuscript.
\end{acknowledgments}


\end{document}